# Media Manipulations in the Coverage of Events of the Ukrainian Revolution of Dignity: Historical, Linguistic, and Psychological Approaches

Ivan Khoma[0000-0003-4607-7065] and Solomia Fedushko*[0000-0001-7548-5856],

and Zoryana Kunch [0000-0002-8924-7274]

Lviv Polytechnic National University, 79013, Lviv, Ukraine

khomaivan@ukr.net, solomiia.s.fedushko@lpnu.ua,
zorjana.kunch@gmail.com

**Abstract.** This article examines the use of manipulation in the coverage of events of the Ukrainian Revolution of Dignity in the mass media, namely in the content of the online newspaper "Ukrainian Truth" ("Ukrainska pravda"), online newspaper "High Castle" ("Vysokyi Zamok"), and online newspaper "ZIK" during the public protest, namely during the Ukrainian Revolution of Dignity. Contents of these online newspapers the historical, linguistic, and psychological approaches are used. Also media manipulations in the coverage of events of the Ukrainian Revolution of Dignity are studied. Internet resources that cover news are analyzed. Current and most popular Internet resources are identified. The content of online newspapers is analyzed and statistically processed. Internet content of newspapers by the level of significance of data (very significant data, significant data and insignificant data) is classified. The algorithm of detection of the media manipulations in the highlighting the course of the Ukrainian revolutions based on historical, linguistic, and psychological approaches is designed. Methods of counteracting information attacks in online newspapers are developed.

## 1 Introduction

Today, the television and print media of the information society have receded into the background. The most popular source of dissemination of information on social and political topics is the channels of mass media distribution are media channels or mass media.

Today, of all the mass media, it is a relevant source of information, namely online media, web services are these are web forums, social networks, online newspapers and others. Since media information is characterized by:

- relevance (interest for the target audience),
- efficiency, versatility (should be interesting for all or as many members of the target audience),
- relative anonymity (behind the media material is the authority not only of the author but also the media) and regularity of influence.

Manipulation in the information sphere is an attempt to use false or biased information to influence the behavior and attitudes of the audience.

Therefore, is important for Ukrainian society courses for people with media literacy is part of media education dedicated condition of critical thinking and understanding attitude towards the media.

The object of study is to investigate the use of manipulation in the coverage of events of the Ukrainian Revolution of Dignity in the mass media.

The subject of the study is the information content on the importance of information of the online newspaper "Ukrainian Truth" ("Ukrainska pravda"), online newspaper "High Castle" ("Vysokyi Zamok"), and online newspaper "ZIK".

The aim of the work is to thoroughly analyze the content, identify techniques for information manipulation in the content of such online newspaper "Ukrainian Truth" ("Ukrainska pravda"), online newspaper "High Castle" ("Vysokyi Zamok"), and online newspaper "ZIK" in period of public protest, namely during the Ukrainian Revolution of Dignity.

The main objectives of the study media manipulations in the coverage of events of the Ukrainian Revolution of Dignity:

- to analyze modern research of international and national scientists;
- to analyze Internet resources that cover news;
- to identify current and most popular Internet resources.
- to analyze the content and statistically process it;
- to classify the content of online newspapers by the level of significance of data (very significant data, significant data and insignificant data);
- to develop the algorithm of detection of the media manipulations in the highlighting the course of the Ukrainian revolutions based on historical, linguistic, and psychological approaches;
- to develop methods of counteracting information attacks in online newspapers.

## 2   Related works

The theme of revolutions is very relevant in the research of world scientists. It should be noted that every year this theme is becoming more popular. This is evidenced by the statistics of the abstract and citation database Scopus. Thus, the result of the search query "revolution" is 36,542 documents with selected year range to analyze from 1823 year to 2020 year.

A graphical representation of the result of the search query "revolution" in the online database Scopus is the chart in Figure 1.

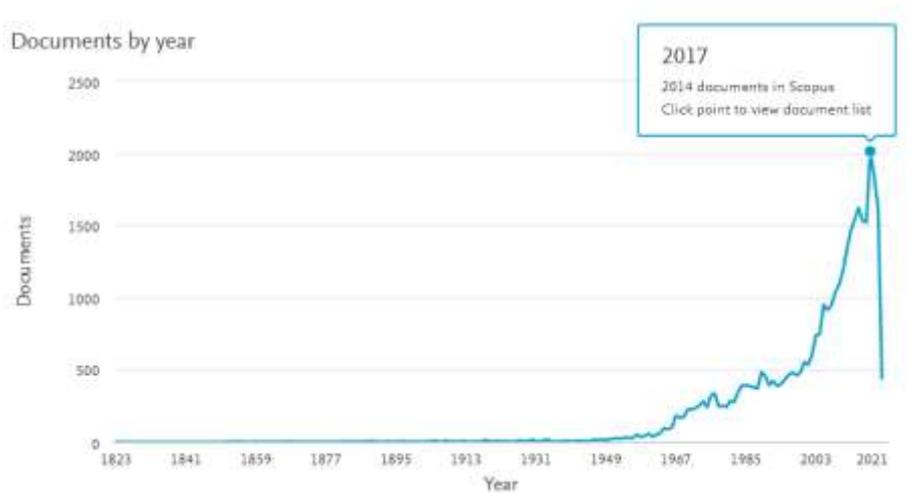

**Fig. 1.** Graphical representation of the result of the search query "revolution" in the abstract and citation database Scopus

As we can see in the chart of figure 1, users of the abstract and citation database Scopus in 2017 year made the most searches query "revolution". These are 2014 documents that cover the events of the revolution.

Scientists are also actively researching historical events in Ukraine. The result of the search query "Ukrainian Revolution" is 205 documents with selected year range to analyze from 1977 year to 2020 year.

A graphical representation of the result of the search query "Ukrainian Revolution" in the abstract and citation database Scopus is the chart in Figure 2.

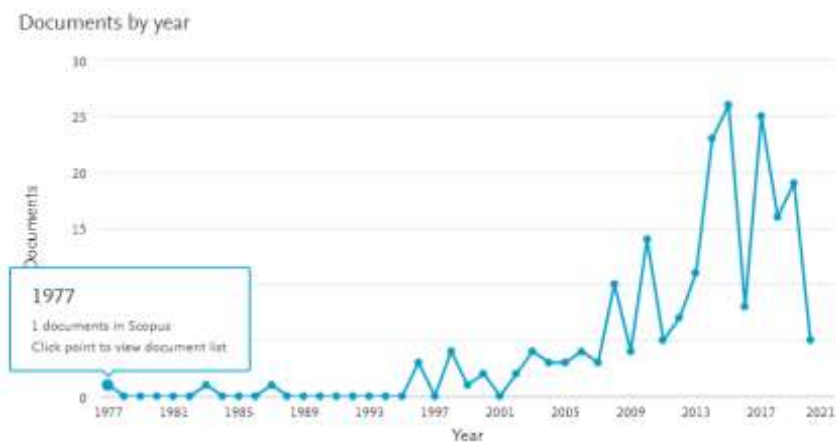

**Fig. 2.** Graphical representation of the result of the search query "Ukrainian Revolution" in the abstract and citation database Scopus

The largest number of documents about the Ukrainian revolutions was published in 2015 year. This is 26 documents, in 2017 year was published in 25 documents, and in 2014 year was published in 23 documents. In 2019 year scientists published 19 works, in 2018 year they published the 16 documents.

The statistics comparing the document counts in Scopus by country or territory of the result of the search query "Ukrainian Revolution" are presented in the chart is shown in Figure 3.

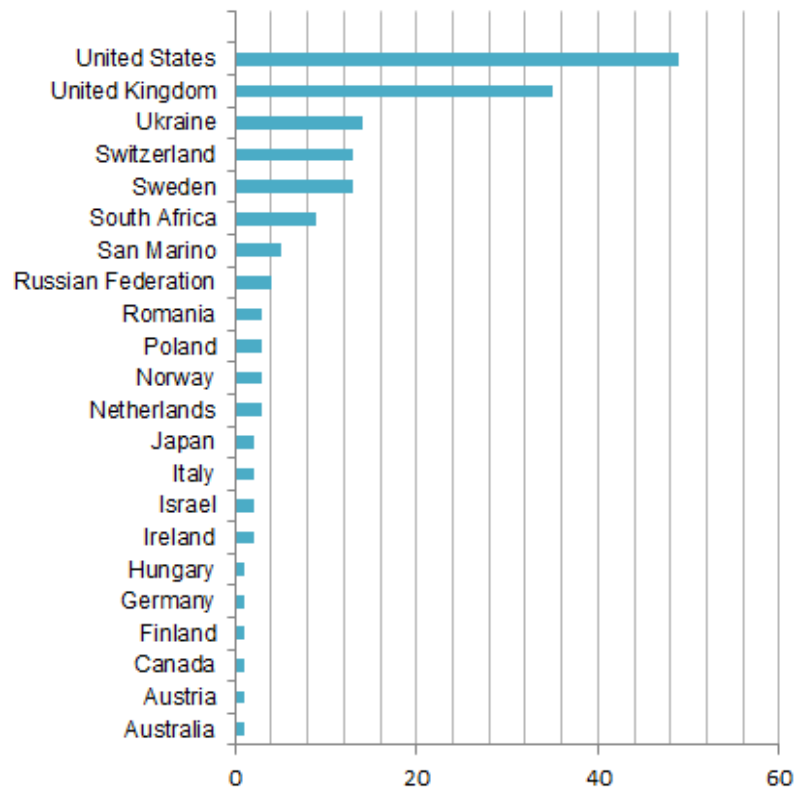

**Fig. 3.** Chart of statistics comparing the document counts in Scopus by country or territory of the search result "Ukrainian Revolution"

As we can see from Figure 3, the largest number of documents about the Ukrainian revolutions in the abstract and citation database Scopus has been published in USA scientific journals is 49 documents. 14 documents on Ukrainian revolutions have been published in Ukrainian publications indexed in the abstract and citation database Scopus.

The search query "ukrainian Revolution of Dignity" is the result of 77 documents with selected year range to analyze from 2015 year to 2020 year.

A graphical representation of the result of the search query "ukrainian Revolution of Dignity" in the abstract and citation database Scopus is the chart in Figure 4.

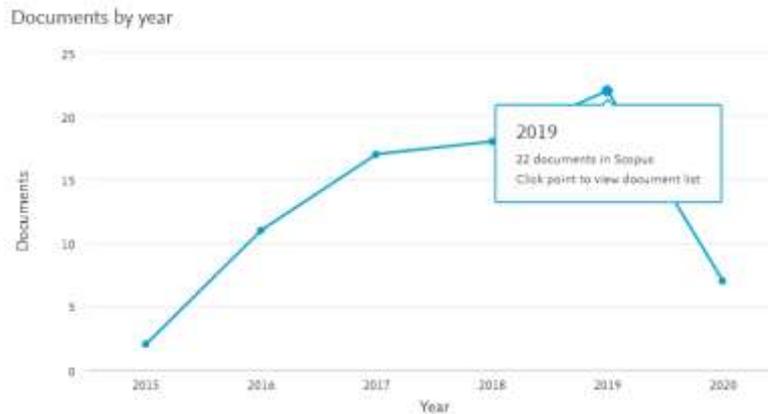

**Fig. 4.** Graphical representation of the search result "ukrainian Revolution of Dignity" in the abstract and citation database Scopus

The largest number of documents about the Ukrainian Revolution of Dignity were published in 2019 year are 22 documents, in 2018 year are published the 18 documents, and in 2017 year are published the 17 documents. In 2016, scientists published 11 papers. It is also worth noting that in a few months in 2020 year, 7 documents on the Ukrainian Revolution of Dignity have already been published in the scientometric database Scopus. The statistics comparing the document counts in Scopus by country or territory of the result of the search query "ukrainian Revolution of Dignity" are presented in the graph shown in Figure 5.

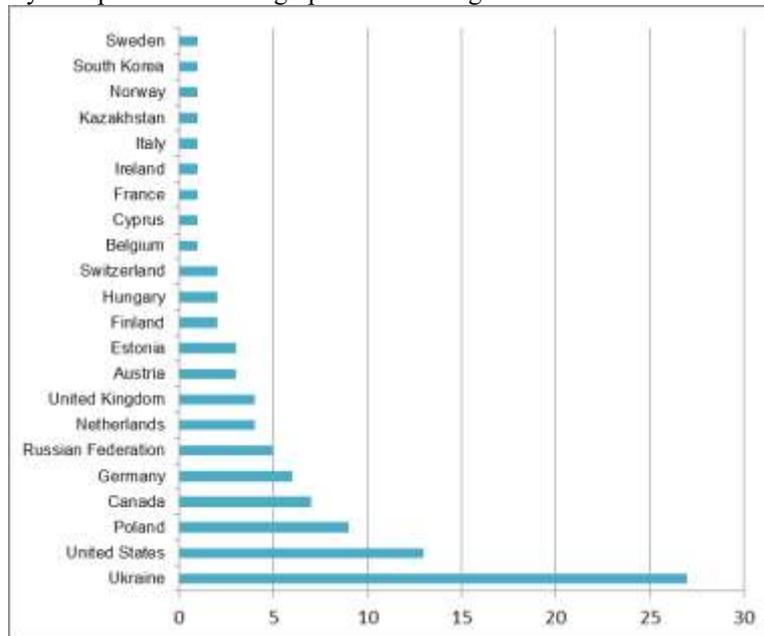

**Fig. 5.** Graph of statistics comparing the document counts in Scopus by country or territory of the search result "ukrainian Revolution of Dignity"

As we can see from Figure 5, the largest number of scientific documents about the Ukrainian Revolution of Dignity have been published in scientific journals of Ukraine in the database Scopus – more than 20 documents. Many documents have also been published in scientific journals in the United States, Poland and Canada.

## 3    Historical approaches in the analysis of media manipulations of the highlighting the course of the Ukrainian Revolution of Dignity

### 3.1    Prerequisites and progress of the revolution.

9 years after the Orange Revolution in Ukraine, at the end of November 2013, the third revolution, dubbed Euromaidan, began. In mid-January 2014, it turned into the Revolution of Dignity, which formally ended in late February 2014. Although in the actions and hearts of many participants in the revolution, it continues to this day.

In February 2010, Viktor Yanukovych was elected President of Ukraine, who through the Constitutional Court managed to return the form of government from parliamentary to presidential. V. Yanukovych's presidency is a time of unconcealed rapprochement of Ukraine with Russia. Ukraine's open preparation for building a union with Russia has been hidden by constant statements about Ukraine's integration with the European Union [1].

The European Union appointed V. Yanukovych to take the exam on November 28, 2013 in Vilnius, Lithuania. The Eastern Partnership Summit was to start this day in Vilnius, where Ukraine together with Georgia and Moldova were to sign an association with the European Union.

However, on November 21, Prime Minister of Ukraine M. Azarov stated that Ukraine is ceasing its preparations for signing the association with the EU. This decision was made "with a view to taking measures to ensure national security, more detailed study and development of a set of measures to be taken to restore lost production volumes and directions of trade and economic relations with the Russian Federation and other members of the UIS". The decision also referred to the resumption of the dialogue "… with the Russian Federation and other Customs Union countries and CIS member states on the revitalization of trade and economic ties in order to preserve and strengthen the joint economic potential of the state"[2].

This government decision was the start of Euromaidan. V. Yanukovych's trip to Vilnius took place at a time when a student protest against the refusal of Ukraine to sign the association with the EU took place in the Independence Square on November 21. Protests swept Lviv, Ivano-Frankivsk, Ternopil and other cities.

Yanukovych did not sign the EU-Ukraine Association at the Eastern Partnership Summit in Vilnius. On the night of November 30, a special unit of the Ministry of Internal Affairs of Ukraine "Golden eagle" beat and dispersed students protesting on Independence Square. As early as December 1, 2013, up to 500,000 protesters gathered on Independence Square and dozens of tents were laid out. A week later, a million protesters gathered for the next rally to identify the goals and objectives of the revolution.

The government headed by V. Yanukovych, organized Antimaidan attracting people from the eastern regions. However, supporters of power have become a marginal phenomenon.

The revolutionary process was led by the leaders of parliamentary opposition V. Klitschko, A. Yatsenyuk, P. Poroshenko and O. Tyagnibok.

Gradually, the protest area also covered the government quarter of Kyiv. At the same time the National Guard and the Berkut Special Forces were brought to Kyiv from all over Ukraine.

The turning point of the Euromaidan was January 16, 2014, when the parliamentary majority, largely represented by the Party of Regions and the Communist Party of Ukraine, passed laws restricting citizens' constitutional rights and freedoms by hand, and with the suspension of the association with the EU threatened the Ukrainian state. These so-called "dictatorial laws" translated protests over Ukraine's EU integration into the Revolution of Dignity. The intentions of the power structures to disperse Euromaidan have escalated into violent confrontations.

On January 22, 2014, the first participants of the Euromaidan - the Dignity Revolution - were killed by firearms. In the western and central regions, participants in the revolution began to seize state power.

V. Yanukovych and his entourage expected that time, frost, insecurity, constant pressure of the security forces would exhaust the moral and physical protesters. This will allow the involvement of special police units in the crackdown on protesters.

On February 18, large-scale confrontations began between protesters and security forces. In the evening Special Forces began storming the square. During the day, 23 members of the Dignity Revolution were killed. By morning, the Maidan managed to hold on. During February 19-21, 57 members of the Dignity Revolution were killed or seriously injured during the standoff. Overall, as of April 11, 2014, 105 members of the Dignity Revolution were killed or died from wounds and injuries, and 102 were in treatment [3]. Generally, the victims of this revolution have been called the "Heavenly Hundred".

On the first anniversary of the Revolution of Dignity, President of Ukraine Petro Poroshenko posthumously awarded the Gold Star "Hero of Ukraine" to almost everyone who was killed during the Revolution of Dignity. This obliged the security forces to return to their locations. The next day, Viktor Yanukovych signed an agreement with the opposition to resolve the crisis in Ukraine. However, he never started to accomplish this task. On the evening of February 21, V. Yanukovych fled to Russia, and on February 22, the Verkhovna Rada adopted a resolution "On the self-removal of the President of Ukraine from the exercise of constitutional powers and the appointment of snap elections of the President of Ukraine".

Russia reacted very cruelly to the events in Ukraine. The annexation of the Autonomous Republic of Crimea will begin in mid-February 2014, and in March 2014 the occupation of Donetsk and Lugansk regions will take place. Russia will start a war against Ukraine.

**Consequences:**

− Ukraine has established itself as a sovereign and independent state;
− Ukraine has once again stopped Russia's intention to return it to its political, economic and cultural zone of influence and definition of development;
− Ukraine has once again confirmed that it shares the European values of development of society and the state and is ready to fight for them.

## 4 The Results of Linguistic and Psychological approaches in the analysis of media manipulations of the highlighting the course of the Ukrainian Revolution of Dignity

Manipulation in the media is a technique of purposeful distortion of information in order to form a certain view, a certain attitude to a particular problem, person and phenomenon. Examples of manipulation in the media can be:

− publication of false data;
− propaganda (white, gray, black);
− expressive language of cruelty;
− providing incomplete information;
− deliberately concealing certain aspects of information;
− shift of accents in the message;
− pulling out of context, etc.

Manipulation in the media is closely linked to propaganda. Propaganda is a form of communication that aims to influence society's attitude to a particular problem, situation and phenomenon. Propaganda is possible through the use of manipulative techniques. Propaganda is purposeful. That is, it is not an accidental mistake or inaccuracy, but a purposeful tactic. Propaganda usually influences attitudes toward certain phenomena or groups of people. For example, propaganda can create hostility towards migrants by portraying them as a threat. Propaganda can take many forms and use different means. Both leaflets distributed on the street and materials in the mass media can be propaganda. This is an additional danger from propaganda - undermining the credibility of the media. Propaganda may be based in part on truthful information, but mixing it with false information leads to actual deception.

Propaganda is a process of spreading facts, views, beliefs to change the attitudes of the individual. At the same time, the communicator informs, explains, persuades, but compromises, manipulates. Positive propaganda is social advertising, PR, which later became widely used. The term propaganda is used in politics.

Prejudice is an antipathy formed towards certain groups of people on the basis of stereotypes. Prejudice creates grounds for discrimination against a group or individual.

Hate speech or hate speech are words and expressions that subconsciously or explicitly program a person for rejection, in particular for aggression against people of other nationalities, religions, life principles, habits. It incites hatred towards a certain group of people on the basis of their common characteristics: nationality, sex, sexual orientation, etc. An important feature of hate speech is that it can be perceived by the communicator as neutral, but it is necessarily offensive and unacceptable to the object of expression.

### 4.1 The algorithm of detection of the media manipulations in the highlighting the course of the Ukrainian revolutions based on historical, linguistic, and psychological approaches

The algorithm of detection of the media manipulations in the highlighting the course The algorithm of detection of the media manipulations in the highlighting of the course of the Ukrainian revolutions based on historical, linguistic, and psychological ap-proaches is developed for definition of manipulations in the Internet media.

This algorithm is shown in Fig. 6.

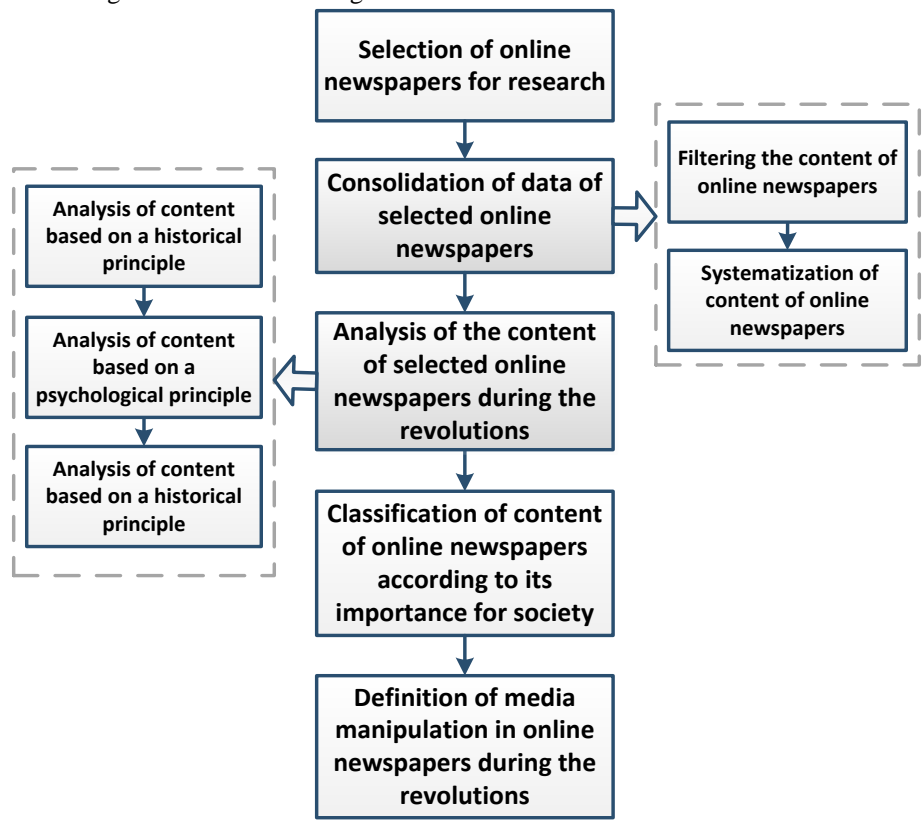

**Fig. 6.** Scheme of algorithm of detection of the media manipulations in the in the coverage of events of the Ukrainian revolutions based on historical, linguistic and psychological approaches

The main stages of algorithm of detection of the media manipulations in the in the coverage of events of the Ukrainian revolutions based on historical, linguistic and psychological approaches are as follows:

1. Selection of online newspapers for research
2. Consolidation of data of selected online newspapers

− Filtering the content of online newspapers
− Systematization of content of online newspapers

3. Analysis of the content of selected online newspapers during the revolutions

− Analysis of content based on a historical principle
− Analysis of content based on a psychological principle
− Analysis of content based on a historical principle

4. Classification of content of online newspapers according to its importance for society
5. Definition of media manipulation in online newspapers during the revolutions

### 4.2 The content analysis of authoritative Ukrainian online newspapers

The contents of three authoritative Ukrainian online newspapers were taken for the content analysis. These online newspapers regularly covered the events of the Revolution of Dignity in Ukraine:

− online newspaper "Ukrainian Truth" ("Ukrainska pravda"),
− online newspaper "High Castle" ("Vysokyi Zamok"),
− online newspaper "ZIK".

Statistics on columns in online newspapers such as online newspaper "Ukrainian Truth" ("Ukrainska pravda"), online newspaper "High Castle" ("Vysokyi Zamok"), and online newspaper "ZIK" were conducted.

After conducting a comprehensive analysis, the study period, from 11/17/2013 to 11/23/2013, was selected. At that time, the most popular columns in online newspapers were "Politika" and "EU".

In the online newspaper "ZIK", in addition to "Politika", the columns "Lviv" and "Western Ukraine" were also popular, as this newspaper is distributed in Western Ukraine.

After investigation of the content of the online newspaper "Ukrainian Truth" ("Ukrainska pravda"), online newspaper "High Castle" ("Vysokyi Zamok"), and online newspaper "ZIK" by the level of significance of data (very significant data, significant data and insignificant data) was analyzed (see Fig. 1).

The created news was also divided into the following categories:

− very significant article;
− significant article;
− insignificant article.

Examples of news according to their level of significance of the online newspaper "Ukrainian Truth" ("Ukrainska pravda"): very significant, significant and insignificant news, presented in Figure 7.

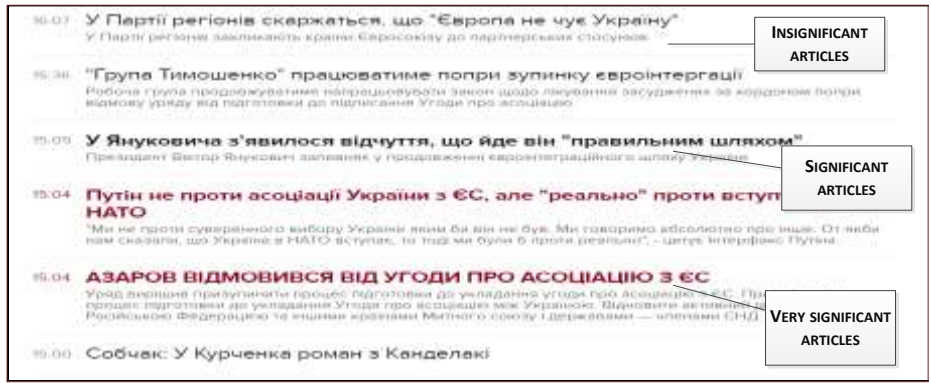

**Fig. 7.** Examples of news according to their level of significance of online newspaper "Ukrainian Truth" ("Ukrainska pravda"): very significant, significant and insignificant news.

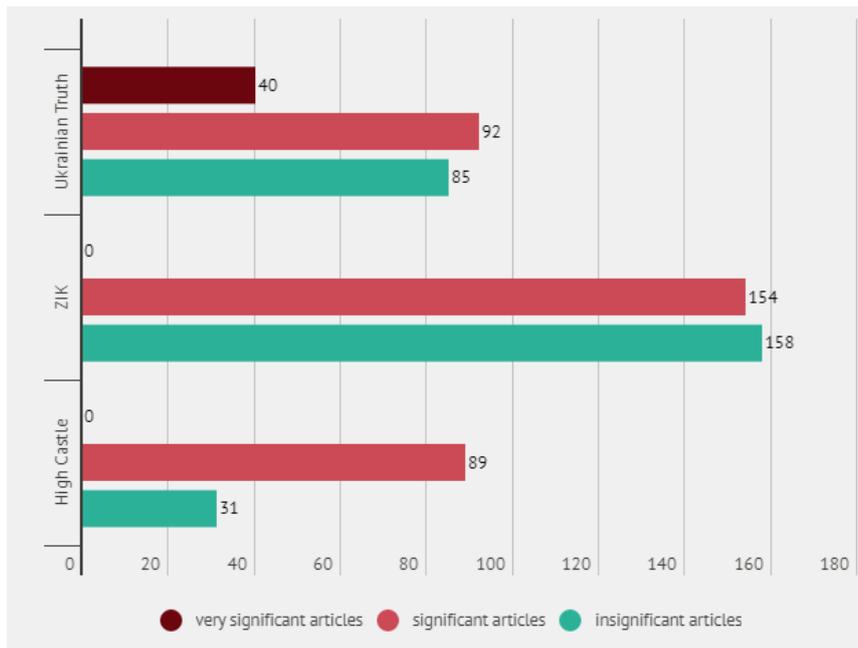

**Fig. 8.** Statistics on the significance of articles

It was establish that in the online newspaper "ZIK" and online newspaper "High Castle" ("Vysokyi Zamok") commenting the articles is impossible. Only in online newspaper "Ukrainian Truth" ("Ukrainska pravda") users commented on articles.

For the study, we chose the online newspaper "Ukrainian Truth" ("Ukrainska pravda"). It was determined that in the period from 11/17/2013 to 11/23/2013, the number of important articles doubled (Fig. 9).

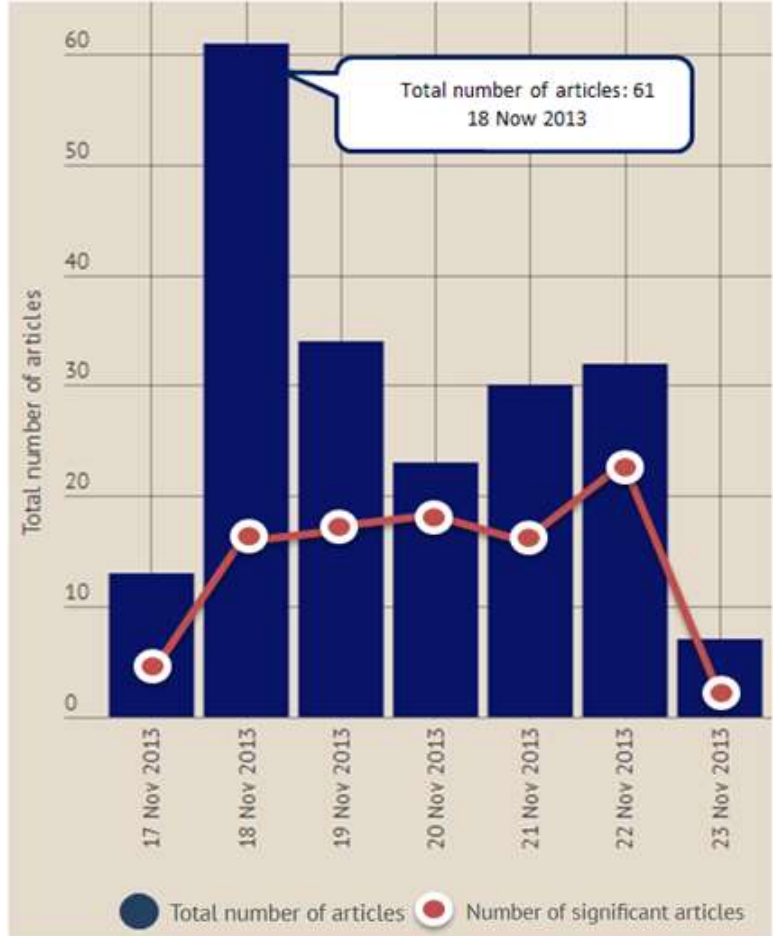

**Fig. 9.** Data on the number of significant articles

This means that during this period there was a fairly intensive attendance of readers and the newspaper carried out active manipulation. That is, they determined the importance of the articles themselves, thus promoting the news that was beneficial to them at the time.

In the online newspaper "Ukrainian Truth" ("Ukrainska pravda") users had various discussions, i.e. they commented and responded to comments. And in these discussions, many conflicts were revealed in which a large number of bots, flames and provocateurs took part. (Fig. 10).

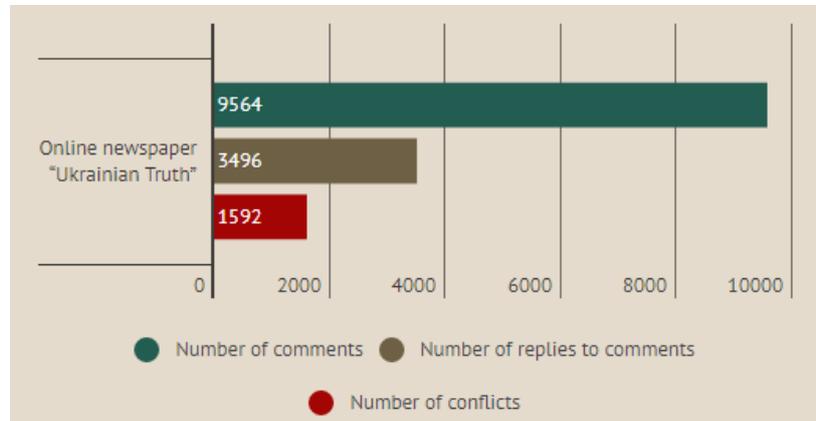

**Fig. 10.** Statistics of commenting in the online newspaper "Ukrainian Truth" ("Ukrainska pravda")

Statistics of commenting in the online newspaper "Ukrainian Truth" ("Ukrainska Pravda") shows that during the Ukrainian Revolution of Dignity it is very important to discuss events in society. This is evidenced by the large number of comments to articles in the online newspaper "Ukrainian Truth" ("Ukrainska Pravda").

## Conclusion

This article examines the use of manipulation in the coverage of events of the Ukrainian Revolution of Dignity in the mass media, namely in the content of the online newspaper "Ukrainian Truth" ("Ukrainska pravda"), online newspaper "High Castle" (Vysokyi Zamok "), And online newspaper" ZIK" during the public protest, namely during the Ukrainian Revolution of Dignity.

Also studied media manipulations in the coverage of events of the Ukraini-an Revolution of Dignity:

— modern researches of international and domestic scientists are analyzed;
— analyzed Internet resources that cover news;
— identified current and most popular Internet resources.
— the content is analyzed and statistically processed;
— classified Internet content of newspapers by the level of significance of data (very significant data, significant data and insignificant data);
— developed the algorithm of detection of the media manipulations in the highlighting the course of the Ukrainian revolutions based on historical, linguistic, and psychological approaches;
— methods of counteracting information attacks in online newspapers have been developed.